\documentclass[acmsmall,screen,authorversion,nonacm]{acmart}
\AtBeginDocument{%
	\providecommand\BibTeX{{%
			Bib\TeX}}}

\usepackage{amsmath,amsfonts}
\usepackage{graphicx}
\usepackage{textcomp}
\usepackage{xcolor}
\usepackage{algpseudocode}
\usepackage{algorithm}
\usepackage{xspace}
\usepackage{listings}
\usepackage{color}
\usepackage{multirow}
\usepackage{dblfloatfix}
\usepackage{lscape} 
\usepackage{xparse}
\usepackage{placeins}
\usepackage{caption} 
\usepackage{subcaption}
\usepackage{enumitem}
\usepackage{array, makecell}
\usepackage{wrapfig}
\usepackage{colortbl}
\usepackage{tabularx} 
\usepackage{booktabs}
\usepackage{courier}
\usepackage{ragged2e}
\usepackage{amsthm}
\usepackage{hyperref}
\usepackage{comment}
\usepackage{pdflscape}
\usepackage{afterpage}
\usepackage{fontawesome}
\usepackage{pifont}
\usepackage{lscape}
\usepackage{enumitem}
\usepackage{fp}
\usepackage{tcolorbox}

\setlist[itemize]{noitemsep, topsep=4pt}
\definecolor{lightgray}{HTML}{f6f6f6}
\definecolor{darkgray}{rgb}{.4,.4,.4}
\definecolor{darkblue}{HTML}{800080}
\definecolor{brickred}{HTML}{b04f4f}
\definecolor{purple}{rgb}{0.65, 0.12, 0.82}
\definecolor{diffadd}{HTML}{288f26}
\definecolor{diffrmbg}{HTML}{ffebe9}
\definecolor{diffaddbg}{HTML}{e6ffeb}
\definecolor{diffremove}{HTML}{de4f54}
\definecolor{carrotorange}{rgb}{0.8, 0.33, 0.0}
\definecolor{highlight}{HTML}{fefbc2}
\definecolor{bluegray}{HTML}{3182bd}

\lstdefinelanguage{JavaScript}{
  keywords={typeof, new, true, false, catch, function, return, null, catch, switch, var, const, let, extends, if, in, while, do, else, case, break, async, await, of},
  keywordstyle=\color{darkblue}\bfseries,
  ndkeywords={class, export, boolean, throw, implements, import, this, setTimeout},
  ndkeywordstyle=\color{brickred}\bfseries,
  identifierstyle=\color{black},
  sensitive=false,
  comment=[l]{//},
  morecomment=[f][\color{diffadd}\bfseries]{+\ },
  morecomment=[s]{/*}{*/},
  morecomment=[f][\color{diffremove}\bfseries]{- },
  commentstyle=\color{violet}\ttfamily,
  stringstyle=\color{carrotorange}\ttfamily,
  morestring=[b]',
  morestring=[b]"
}

\lstdefinelanguage{Python}{
  keywords={typeof, new, true, false, catch, function, return, null, catch, switch, var, const, let, extends, if, in, while, do, else, case, break, async, await, of, from, import, class, def},
  keywordstyle=\color{darkblue}\bfseries,
  ndkeywords={class, export, boolean, throw, implements, import, this, setTimeout, self, __init__},
  ndkeywordstyle=\color{brickred}\bfseries,
  identifierstyle=\color{black},
  sensitive=false,
  comment=[l]{//},
  morecomment=[f][\color{diffadd}\bfseries]{+\ },
  morecomment=[s]{/*}{*/},
  morecomment=[f][\color{diffremove}\bfseries]{- },
  commentstyle=\color{violet}\ttfamily,
  stringstyle=\color{carrotorange}\ttfamily,
  morestring=[b]',
  morestring=[b]"
}

\lstdefinelanguage{json}{
  keywords={<system_prompt>, <user_prompt>},
  otherkeywords={<system_prompt>, <user_prompt>},
  keywordstyle=\color{darkblue}\bfseries,
  ndkeywords={<system_prompt>, <user_prompt>},
  ndkeywordstyle=\color{brickred}\bfseries,
  identifierstyle=\color{black},
  sensitive=false,
  comment=[l]{//},
  morecomment=[f][\color{diffadd}\bfseries]{+\ },
  morecomment=[s]{/*}{*/},
  morecomment=[f][\color{diffremove}\bfseries]{- },
  commentstyle=\color{violet}\ttfamily,
  morestring=[b]',
  morestring=[b]"
}

\theoremstyle{definition}
\newtheorem{definition}{Definition}
\newcommand{\header}[1]{\par\smallskip\noindent\textbf{#1.}}

\def\BibTeX{{\rm B\kern-.05em{\sc i\kern-.025em b}\kern-.08em
    T\kern-.1667em\lower.7ex\hbox{E}\kern-.125emX}}
    
\newboolean{showcomments}
\setboolean{showcomments}{true}
\ifthenelse{\boolean{showcomments}}
{
	\definecolor{myyellow}{RGB}{255, 228, 26}
	\definecolor{myblue}{RGB}{50, 50, 220}
	\newcommand{\nb}[2]{
		{\sf
			\fcolorbox{myyellow}{yellow}{\scriptsize\textbf{#1}}%
			$\blacktriangleright$%
			{\color{myblue}\fontsize{7pt}{8pt}\selectfont\textbf{#2}}%
		}%
	}
}
{
	\newcommand{\nb}[2]{}
}

\newcommand{\toolname}{\textsc{NaviQAte}\xspace}
\newcommand{\example}{\emph{Find a black blazer for men with L size and add to wishlist}\xspace}

\newcommand{\absexample}{\emph{Add a specific type of clothing item in a particular size to a wishlist}\xspace}

\newcommand{\gptthree}{\textsc{GPT-4o mini}\xspace}
\newcommand{\gpt}{\textsc{GPT-4o}\xspace}

\newcommand{\webcanvasconcretesr}{38.46}
\newcommand{\webcanvasconcretetos}{0.39}
\newcommand{\webcanvasabstractsr}{28.84}
\newcommand{\webcanvasabstracttos}{0.16}

\newcommand{\oursconcretesr}{44.23}
\newcommand{\oursconcretetos}{0.58}
\newcommand{\oursabstractsr}{38.46}
\newcommand{\oursabstracttos}{0.56}

\newcommand{\gptconcretesr}{7.69}
\newcommand{\gptconcretetos}{0.02}
\newcommand{\gptabstractsr}{7.69}
\newcommand{\gptabstracttos}{0.02}

\FPeval{\comparisonconcrete}{round((\oursconcretesr-\webcanvasconcretesr)/\webcanvasconcretesr*100, 0)}

\FPeval{\comparisonconcretetos}{round((\oursconcretetos-\webcanvasconcretetos)/\webcanvasconcretetos*100, 0)}

\FPeval{\comparisonabstract}{round((\oursabstractsr-\webcanvasabstractsr)/\webcanvasabstractsr*100, 0)}

\FPeval{\comparisonabstracttos}{round((\oursabstracttos-\webcanvasabstracttos)/\webcanvasabstracttos*100, 0)}

\FPeval{\comparisonconcretegpt}{round((\oursconcretesr-\gptconcretesr)/\gptconcretesr*100, 0)}

\FPeval{\comparisonabstractgpt}{round((\oursabstractsr-\gptabstractsr)/\gptabstractsr*100, 0)}

\makeatletter
\DeclareRobustCommand{\change}{%
  \@bsphack
  \leavevmode
  \color{blue}
  \@esphack
}
\DeclareRobustCommand{\stopchange}{
  \@bsphack
  \normalcolor
  \@esphack
}
\makeatother

\begin{document}

\title{\toolname: Functionality-Guided Web Application Navigation}

% \title{Functionality-Guided Web Application Exploration}

\author{Mobina Shahbandeh}
\orcid{0000-0002-1878-8399}
\affiliation{%
  \institution{University of British Columbia}
  \city{Vancouver}
  \country{Canada}
}
\email{mobinashb@ece.ubc.ca}

\author{Parsa Alian}
\orcid{0009-0002-8335-2724}
\affiliation{%
  \institution{University of British Columbia}
  \city{Vancouver}
  \country{Canada}
}
\email{palian@ece.ubc.ca}

\author{Noor Nashid}
\orcid{0000-0002-9889-2312}
\affiliation{%
  \institution{University of British Columbia}
  \city{Vancouver}
  \country{Canada}
}
\email{nashid@ece.ubc.ca}

\author{Ali Mesbah}
\orcid{0000-0002-8783-7967}
\affiliation{%
  \institution{University of British Columbia}
  \city{Vancouver}
  \country{Canada}
}
\email{amesbah@ece.ubc.ca}

\thispagestyle{plain}
\pagestyle{plain}

\begin{abstract}

End-to-end web testing is challenging due to the need to explore diverse web application functionalities. Current state-of-the-art methods, such as WebCanvas, are not designed for broad functionality exploration; they rely on specific, detailed task descriptions, limiting their adaptability in dynamic web environments. We introduce \toolname, which frames web application exploration as a question-and-answer task, generating action sequences for functionalities without requiring detailed parameters. Our three-phase approach utilizes advanced large language models like \gpt for complex decision-making and cost-effective models, such as \gptthree, for simpler tasks. \toolname focuses on functionality-guided web application navigation, integrating multi-modal inputs such as text and images to enhance contextual understanding. Evaluations on the Mind2Web-Live and Mind2Web-Live-Abstracted datasets show that \toolname achieves a \oursconcretesr\% success rate in user task navigation and a \oursabstractsr\% success rate in functionality navigation, representing a \comparisonconcrete\% and \comparisonabstract\% improvement over WebCanvas. These results underscore the effectiveness of our approach in advancing automated web application testing.

\end{abstract}



\maketitle

% !TEX root =  ../main.tex
\section{Introduction}
In recent years, web applications have become ubiquitous, serving as essential tools for a wide range of online activities, from e-commerce to social networking. With over 781 billion website visits globally each month \cite{xavier2024webunpackedquantitativeanalysis}, their popularity highlights the growing need for developers to maintain high standards of quality and functionality. Traditional manual web testing approaches, however, can be time-consuming and challenging \cite{biagiola2019}, leading to the increased adoption of automated testing methodologies to streamline the quality assurance process~\cite{IMTIAZ2021110841,gur:webagent:iclr2024,lai2024autowebglm,chen:webvln:aaai2024,pan:mind2web-live:arxiv2024,thil2024navigating,yao:webshop:neurips2022,xu2021grounding,li2021glider,zheng2021automatic,alian2024bridging, deng2024mind2web}.
Despite these advances, conventional testing tools may exhibit challenges and shortcomings regarding testing coverage, potentially overlooking critical bugs and usability issues \cite{gur:webagent:iclr2024, furuta2023multimodal}. The discrepancy between tests generated by conventional methods and real user interactions further compounds these challenges \cite{zheng:gpt4v:arxiv24}, resulting in suboptimal testing outcomes.

Web applications typically encompass a spectrum of actions, including form submissions, button clicks, and navigation through various pages. Automated testing tools for web applications encounter challenges stemming from the intricate and dynamic nature of modern web interfaces, which can feature diverse layouts, interactions, and non-deterministic states \cite{adamsan2017}.

To address these challenges and mitigate the limitations of the traditional test generation methods, there has been a growing interest in leveraging deep learning (DL) \cite{chen:webvln:aaai2024,deng2024mind2web} and reinforcement learning (RL) \cite{kontogiannis:tree-based-crawling, lai2024autowebglm,thil2024navigating,li2021glider,Han2018FocusedCT,gur2018learning,liu2018reinforcement:miniwoB:plus,zheng2021automatic} techniques for automated testing in web applications. By assimilating insights from human testers' behavior, such automated testing approaches aim to emulate human-like interactions with web interfaces, thereby improving the comprehensiveness and effectiveness of testing.
However, these DL and RL-based methodologies are not without their constraints. Challenges such as requiring a large corpus of training data, limited generalization capabilities, and the non-deterministic nature of web applications pose hurdles to their efficacy \cite{furuta2023multimodal}.

% In this context, Large Language Models (LLMs)~\cite{zhao:llm-survey:arxiv23, minaee:llm-survey:arxiv24} and Large Multi-Modal Models (LMMs)~\cite{mitra:cot-lmm:cvpr24, fu:llm:arxiv23, fu:lmm-with-video:arxiv24, yin:llm-survey:arxiv23, yin:woodpecker-lmm:arxiv23, fu:challenger-gpt4-gemini-lmm:arxiv23} emerge as a promising avenue for addressing the complexities of automated testing in web applications. LLMs, exemplified by models such as GPT-3, exhibit remarkable proficiency in natural language understanding and can adeptly engage in question-answering tasks. LMMs, exemplified by models such as \gpt, demonstrate a substantial understanding of visual and textual context. Leveraging the capabilities of language models, particularly in comprehending human knowledge and simulating human-like interactions, holds potential for revolutionizing the landscape of automated testing methodologies for web applications.

In this context, Large Language Models (LLMs)~\cite{zhao:llm-survey:arxiv23, minaee:llm-survey:arxiv24,mitra:cot-lmm:cvpr24, fu:llm:arxiv23, fu:lmm-with-video:arxiv24, yin:llm-survey:arxiv23, yin:woodpecker-lmm:arxiv23, fu:challenger-gpt4-gemini-lmm:arxiv23} represent a promising solution for addressing the complexities of automated testing. LLMs, such as \gpt, exhibit a robust understanding of both visual and textual contexts. By leveraging the capabilities of these models—particularly their ability to comprehend human knowledge and simulate human-like interactions—there is significant potential to revolutionize automated testing methodologies for web applications. LLMs have also been employed in the literature across a wide range of autonomous agents~\cite{Wang2023ASO,yan2023gpt4v}, including web navigation agents, where they serve as the decision-making backbone~\cite{ma2024laserllmagentstatespace,pan:mind2web-live:arxiv2024,zheng:gpt4v:arxiv24,lù2024weblinxrealworldwebsitenavigation}.

However, a notable limitation of existing agents is their emphasis on testing specific, isolated tasks, rather than evaluating comprehensive functionalities. In practical applications, testing requirements frequently encompass broader functionalities such as ``Search for an item'' or ``Add an item to the cart.'' These functionalities represent more generalized actions compared to the highly specific tasks (e.g., searching for a particular item or adding a specific item to the cart) that many current agents are designed to address. This gap underscores the necessity for a more holistic approach to web application testing. 

% However, these existing agents focus on testing specific tasks rather than entire functionalities. In practice, we often need to test a whole functionality rather than isolated tasks.
% \nashid{Can we clarify what is meant by “functionality-aware” and how this differentiates from isolated "task-based" or "action-based" approaches?}

%\change To address the aforementioned issues, we propose a novel approach to web application testing by framing it as a question-and-answer (Q\&A) task. Our technique, called \toolname, leverages the advanced reasoning capabilities of LLMs to simulate human testers, aiming to improve the efficiency and effectiveness of automated web testing. \toolname incorporates multiple context extraction processes, an option scoring system, and an action grounding mechanism to ensure accurate action selection and execution throughout the exploration process. \toolname focuses on exploring functionalities and eliminates the need for predefined tasks and parameters, facilitating the automated web application testing process. It generates parameters and steps dynamically and selects the best action from a comprehensive list of options in every step.\stopchange

To address the aforementioned challenges, we propose a novel approach, \toolname, for web application navigation, structured around three key phases: action planning, choice extraction, and decision making.

First, \toolname generates concrete functionality descriptions using retrieval-augmented generation to transform abstract tasks into actionable task descriptions. From these descriptions, we extract webpage context from multi-modal sources, such as meta tags, prior actions, and screenshots, to create an abstract representation of the current state, thereby enhancing the model’s understanding of the web application's structure and functionality. This context is leveraged to predict the next step. We then identify actionable elements on the webpage and refine them using a novel \emph{choice ranking} system. Elements are ranked based on their semantic similarity to the predicted next step, with additional context provided by nearby visual and textual information. This ensures that only the most relevant and contextually appropriate elements are considered for action, avoiding redundant decisions and improving accuracy and efficiency. Finally, \toolname selects the optimal action by combining task history, annotated screenshots, and ranked actionable elements. Visual cues help the model interpret the spatial layout and context, allowing for accurate and efficient navigation. This dynamic adaptation ensures precise task execution as \toolname executes through each step of the task. 

In this paper, we make the following contributions:

\begin{itemize}
\item A novel functionality-guided approach to web application exploration, framing web navigation as a question-and-answer task that dynamically transforms abstract functionality descriptions into concrete task steps.

\item A multi-phase, multi-model methodology that enhances web navigation by integrating action planning, choice extraction, and decision making, leveraging LLMs to process textual and visual inputs for greater contextual accuracy.

\item To the best of our knowledge, we are the first to enable functionality-guided navigation for automated web exploration, moving beyond predefined task-based methods. 

\end{itemize}

Our empirical evaluations demonstrate that \toolname attains success rates of \oursconcretesr\% and \oursabstractsr\% in the execution of user tasks and functionalities, respectively. Notably, these results represent a substantial improvement over the next-best baseline, WebCanvas \cite{pan:mind2web-live:arxiv2024}, highlighting improvements of \comparisonconcrete\% and \comparisonabstract\% in task and functionality success rates, respectively.

\section{Motivation}
\label{sec:motivation}

In end-to-end (E2E) web testing, the primary goal is often to validate the functionality of an entire system as a whole \cite{LEOTTA2016193}, rather than focusing on individual, concrete tasks. Concrete tasks require detailed parameters and predefined steps, which becomes impractical and unsustainable as applications grow in complexity. Defining these specific tasks for every possible interaction is not scalable. In contrast, focusing on functionalities allows testers to assess the overall behavior of the application without needing to define every detail. Prioritizing functionality over predefined tasks enhances the adaptability, scalability, and coverage of the testing process, leading to more reliable and robust web applications.

\begin{figure*}[h]
    \centering

    \begin{subfigure}[b]{0.32\textwidth}
         \centering
         \includegraphics[width=0.985\textwidth]{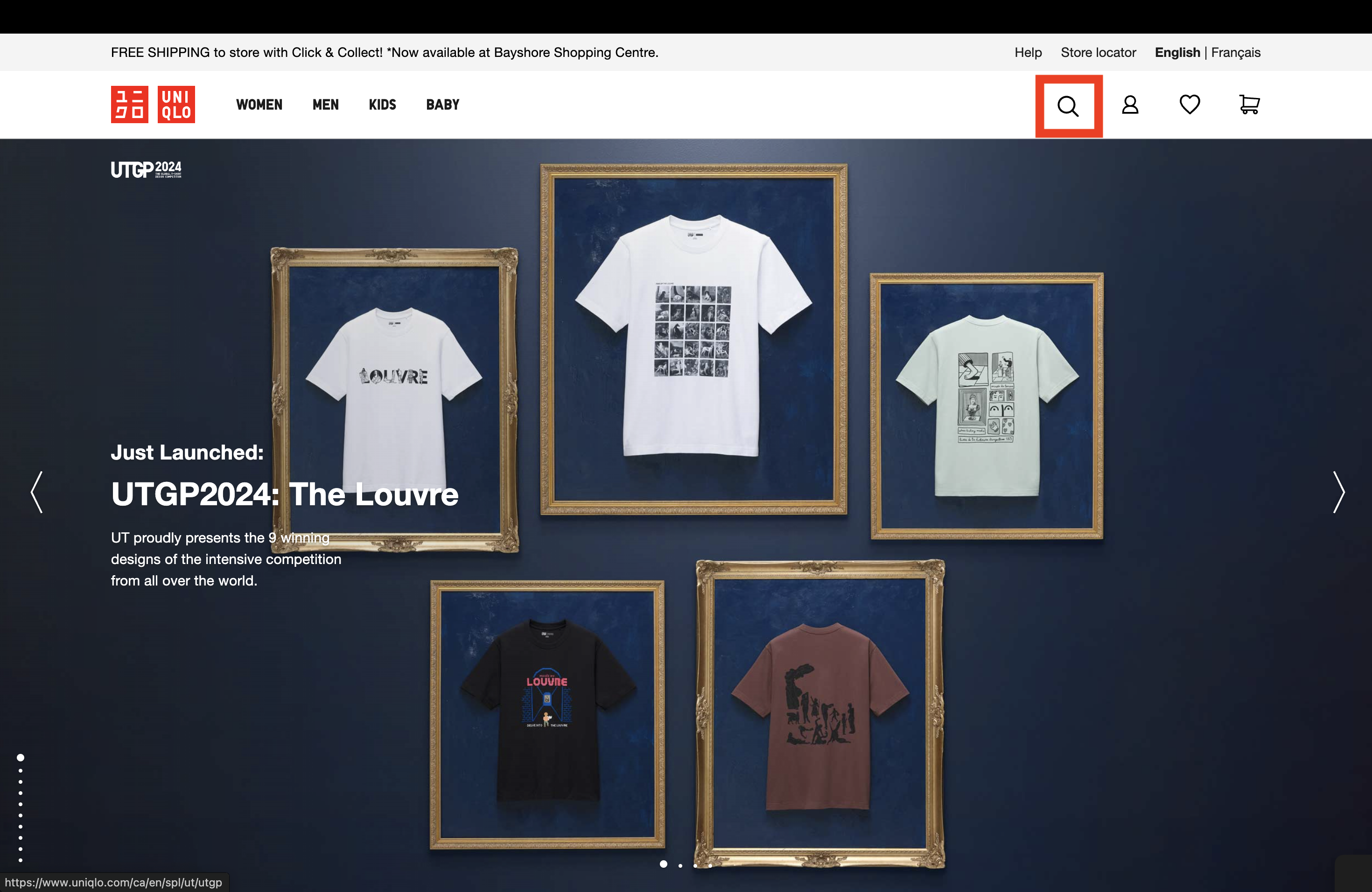}
         \caption{Select search button}
         \label{fig:sample-task-1}
     \end{subfigure}
     \begin{subfigure}[b]{0.32\textwidth}
         \centering
         \includegraphics[width=0.985\textwidth]{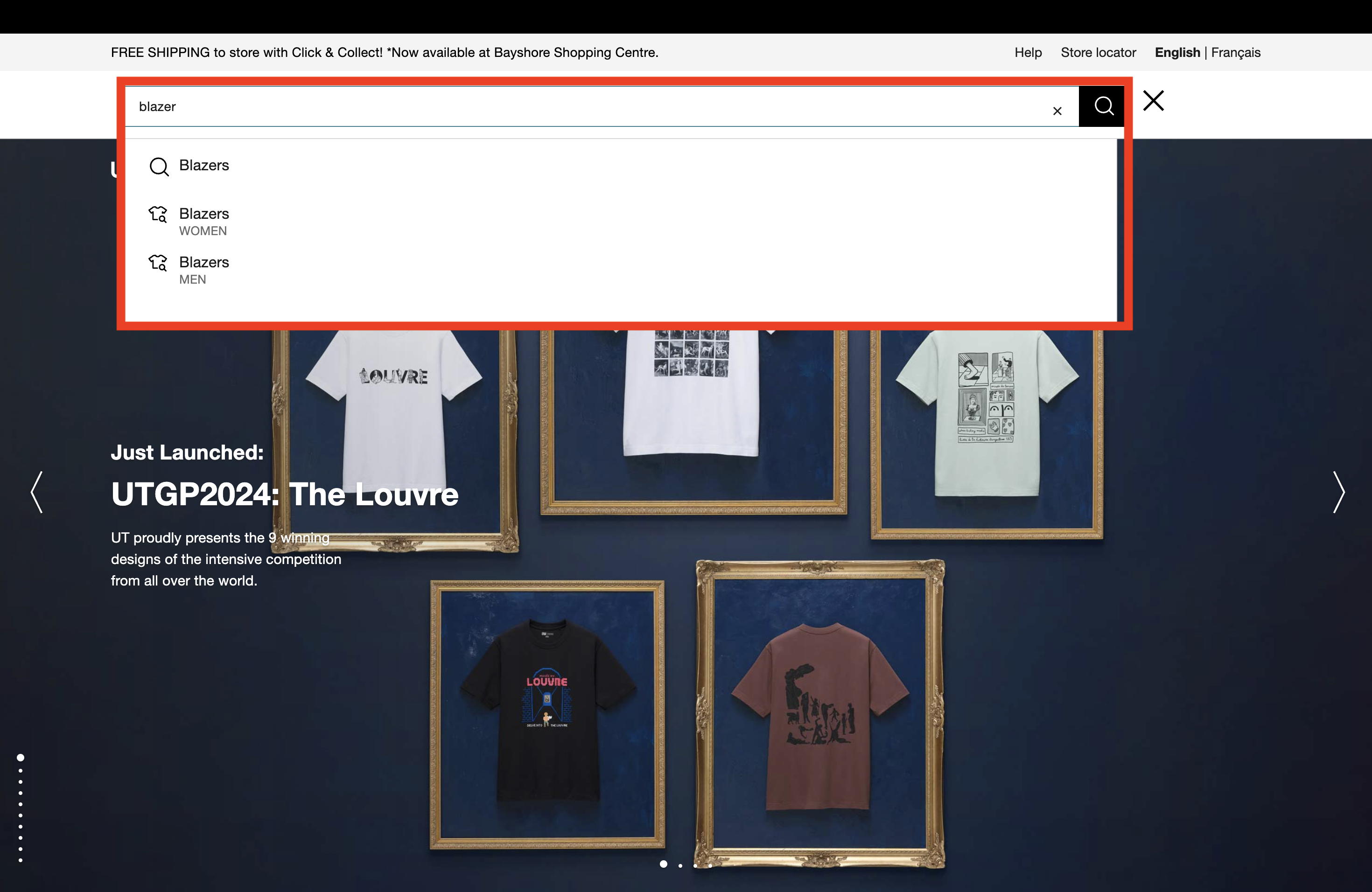}
         \caption{Input search value}
         \label{fig:sample-task-2}
     \end{subfigure}
     \begin{subfigure}[b]{0.32\textwidth}
         \centering
         \includegraphics[width=0.985\textwidth]{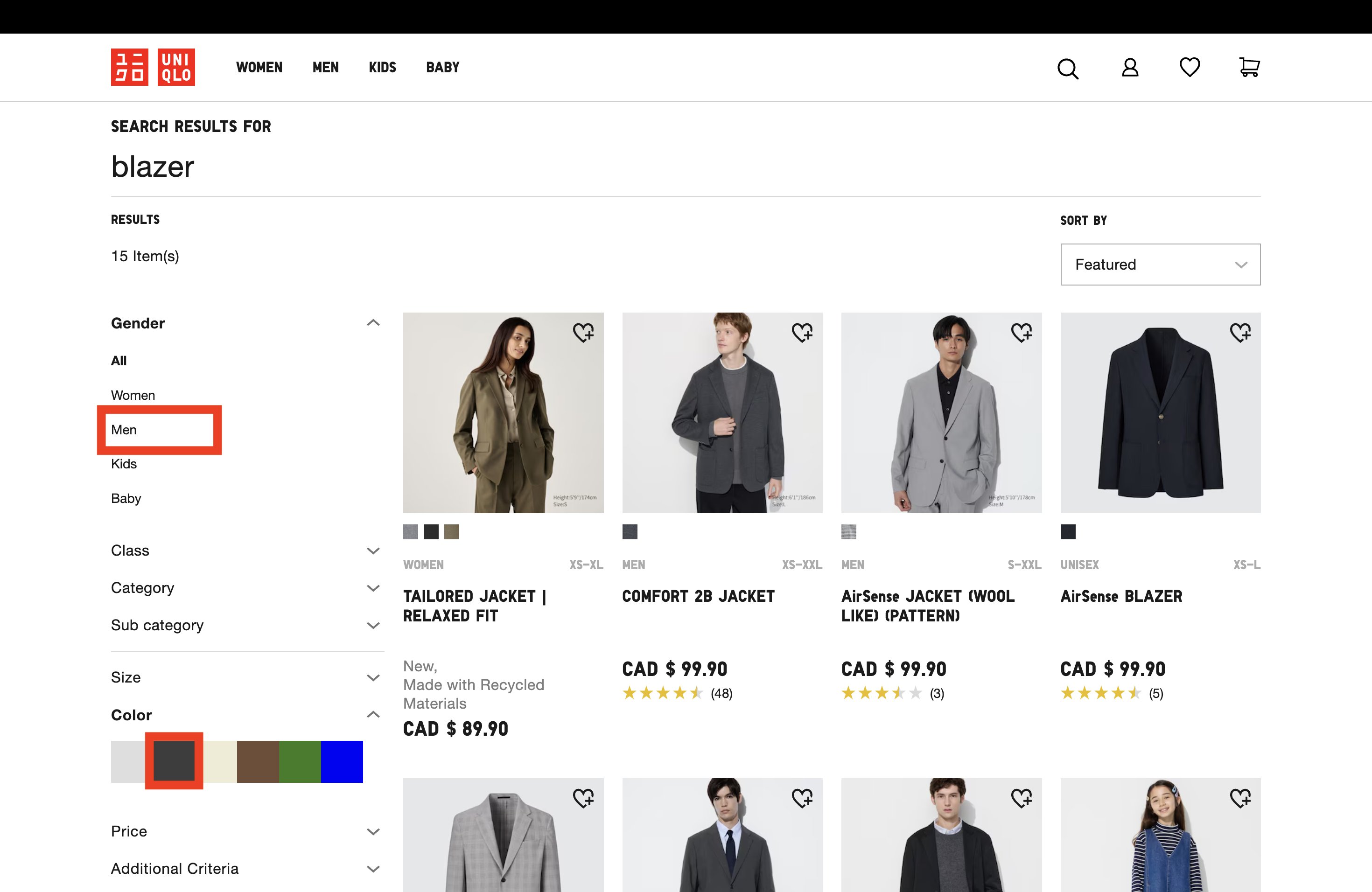}
         \caption{Select filters}
         \label{fig:sample-task-3}
     \end{subfigure}

    \vspace{1em}

     \begin{subfigure}[b]{0.32\textwidth}
         \centering
         \includegraphics[width=0.985\textwidth]{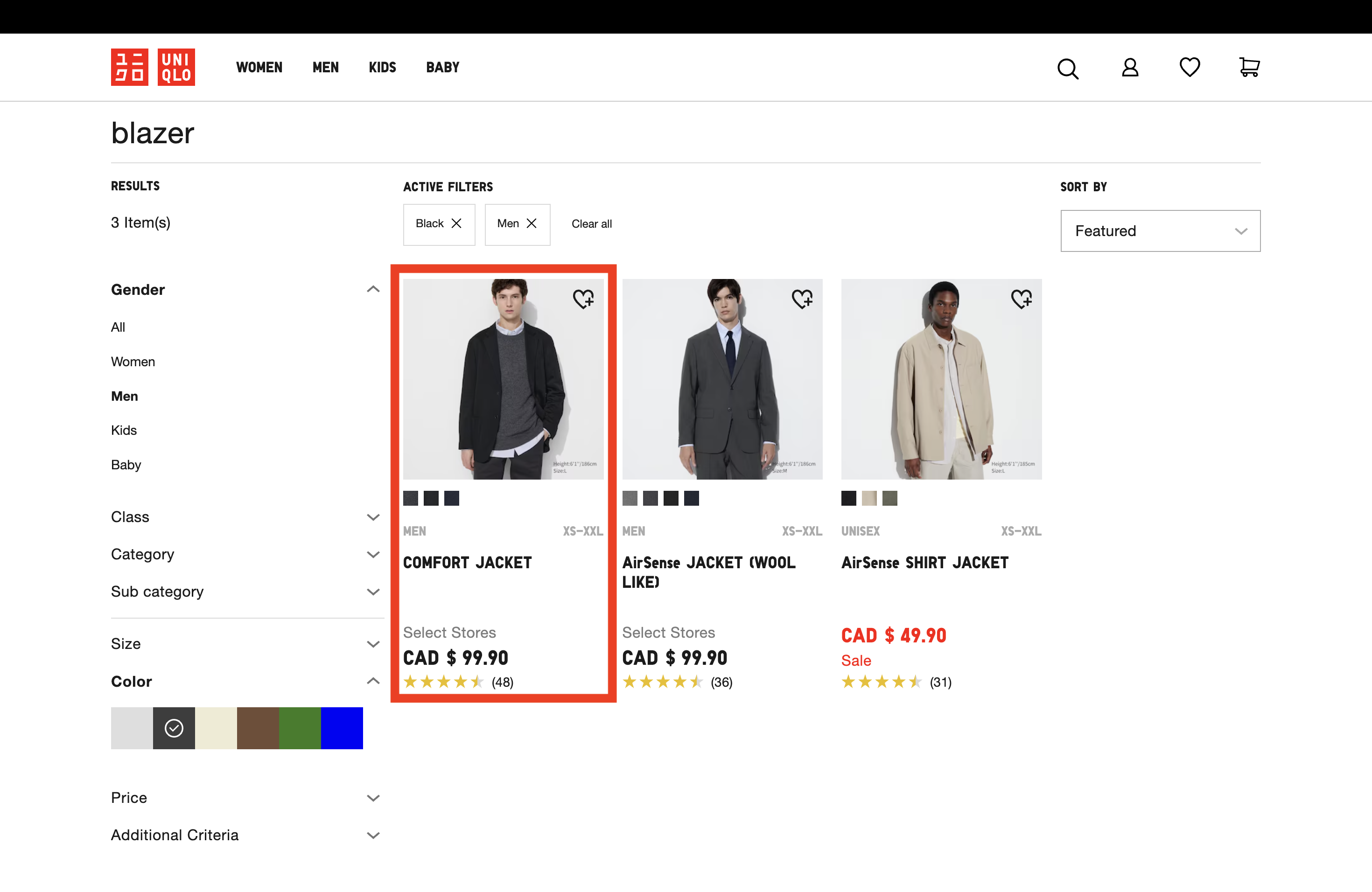}
         \caption{Select product}
         \label{fig:sample-task-4}
     \end{subfigure}
     \begin{subfigure}[b]{0.32\textwidth}
         \centering
         \includegraphics[width=0.985\textwidth]{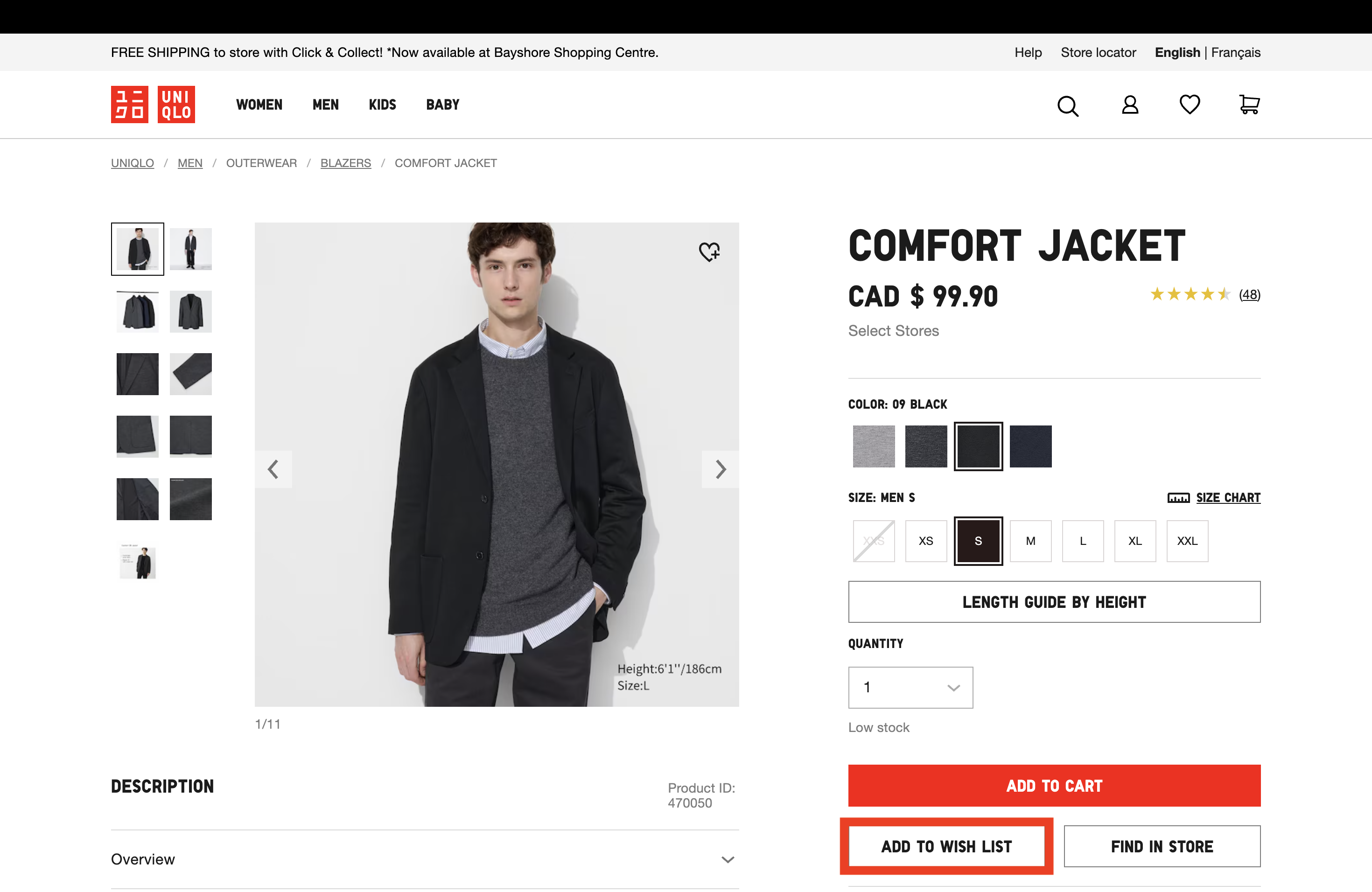}
         \caption{Add to wishlist}
         \label{fig:sample-task-4}
     \end{subfigure}
    
    \caption{Sample task execution steps for ``\example''}
    \label{fig:sample-task}
\end{figure*}

To test a web application, developers often define user tasks or scenarios and execute those tasks to verify intended functionalities. An example of such tasks is shown in Figure \ref{fig:sample-task}, where the task being performed is the following: \example. The sequence of actions a user might follow to complete this task could be: 
1) Select the \emph{search} button at the top of the page, 2) Enter the value \emph{Blazer} into the search bar, 3) Apply filters for the \emph{Men} category, \emph{L} size, and \emph{Black} color, 4) Select one of the products, and 5) Click on the \emph{Add to Wishlist} button.

To clarify key terms, we define the following concepts:

\begin{definition}[User Task] 
\label{def:user-task}
A \emph{user task} is a detailed description of a specific operation a user can perform on a web application. This includes any specific attributes or inputs required to execute the operation. A user task consists of a finite sequence of user actions leading to the intended outcome.

Each defined task can be decomposed into two components:
\begin{itemize}[leftmargin=*]
    \item \textbf{Functionality ($\mathcal{F}$):} An abstract description of the task, without specific details.
    \item \textbf{Task parameters ($\mathcal{P}$):} A set of values required to execute the specific task, including inputs, attributes, filters, or conditions.
\end{itemize}
\end{definition}

Each functionality $\mathcal{F}$ represents a broad group of specific user tasks. Rather than focusing on detailed parameters, a functionality provides an abstract view of the intended action.

For example, a motivating task could consist of:
\begin{itemize}
    \item Adding a specific item of a specific size to the wishlist as the functionality ($\mathcal{F}$).
    \item \textit{Blazer}, \textit{Men}, \textit{L}, and \textit{Black} as the task parameters ($\mathcal{P}$).
\end{itemize}

Each task defined this way can be executed through numerous variations. For instance, the task might be completed by navigating to the men's clothing category, selecting the blazer category, applying filters, and then adding one of the resulting products to the wishlist. Alternatively, the user might search for a blazer, apply filters, and then add one of the resulting products to the wishlist.

There are several advantages to exploring functionalities, such as \absexample, over specific tasks, such as \example. By understanding how to execute a functionality, the tool can gain the flexibility to perform any specific task related to that functionality, without needing to learn each task individually. This flexibility is especially valuable in applications such as virtual assistants or E2E testing, where developers only need to reference available features rather than specifying detailed \emph{task parameters} for the agent to operate. As a result, task automation becomes more streamlined. Furthermore, because web applications frequently evolve, predefined user tasks often become obsolete over time, requiring regular updates \cite{pan:mind2web-live:arxiv2024}. Focusing on functionalities rather than concrete tasks offers greater efficiency, as invoking the same functionality in multiple test cases—such as login—often leads to repeated code fragments across all test scripts \cite{LEOTTA2016193}.

\header{Limitations of Existing Methods} 
Recent research has seen increasing interest in automating web tasks, such as web question answering~\cite{webqa, webqaprogramsynth}. Efforts have led to the development of natural language (NLP) models, vision models \cite{chen:webvln:aaai2024}, and reinforcement learning models \cite{li2021glider, liu2018reinforcement:miniwoB:plus, zheng2021automatic}. However, training these task-specific models is resource-intensive and time-consuming. Moreover, they often exhibit suboptimal performance when encountering novel application domains not seen during training.

In contrast, more recent approaches leverage LLMs for web task automation~\cite{gur:webagent:iclr2024, lai2024autowebglm, chen:webvln:aaai2024, pan:mind2web-live:arxiv2024, thil2024navigating, yao:webshop:neurips2022, alian2024bridging}. Most of these methods focus on workflows that employ LLMs to execute specific tasks. However, as previously noted, focusing on task abstractions offers significant advantages over individual tasks. To our knowledge, no prior work has explicitly pursued this direction.

Key challenges remain in developing these systems: (a) Generalization: handling variations in tasks, such as different product types or web layouts, (b) Context Awareness: the agent must understand the entire flow of actions rather than treating steps in isolation, (c) Dynamic Environments: web pages frequently change states (e.g., after applying a filter), requiring the agent to adapt accordingly, and (d) Holistic Testing: beyond one-off task execution, the system must evaluate the entire functionality (e.g., completing an entire shopping flow).

These challenges underscore the need for a more advanced, functionality-driven approach to web navigation and testing. Context awareness is crucial, as systems must make informed decisions that align with overall task objectives rather than isolated steps. In dynamic environments, where web applications and their data are constantly evolving, predefined tasks quickly become outdated and ineffective. Additionally, the focus on holistic testing is essential for evaluating entire functionalities rather than isolated tasks, ensuring a more comprehensive and robust testing strategy.
\section{Approach}
\label{sec:approach}

\subsection{Problem Formulation}

We formulate the web application navigation problem as a \textit{sequential decision-making process} where the goal is to navigate through a series of web states to complete a given task. Each step involves selecting an appropriate action from a predefined action space, based on the current state of the web page, the task at hand, and the predicted next step toward task completion.

\toolname is tasked with generating a sequence of executable actions $A = [A_1, A_2, \dots, A_n]$ to fulfill a given task $U$. At each state $S_i$, the model generates an action $A_i$ by considering the current state, the next predicted step $N_i$, and the overall task objective $U$. Each action $A_i$ consists of three components: the actionable element $(a_i)$, the action type $(t_i)$ (e.g., \emph{click}, \emph{type}, or \emph{select}), and the input (if applicable) $(i_i)$.

\begin{equation}
    A_i = f(S_i, U, N_i) = (a_i, t_i, i_i)
\end{equation}

\subsection{Actionable Elements}
\label{subsec:actionables}
We focus on a variety of actionable elements, including navigation links, buttons (e.g., ``Submit,'' ``Next''), input fields (e.g., search boxes), select elements, and any other elements that trigger actions (e.g., elements with event listeners). 

The state $S_i$ at time step $i$ is represented by the set of all available actionable elements $[a_0, a_1, ..., a_n]$ in that state, and $n$ represents the total number of actionable elements in that state, as well as all the actions performed before reaching this state $[A_1, A_2, ..., A_{i-1}]$.

\subsection{Action Space}
\label{subsec:actionspace}
At any given state $S_i$, the available actions are defined as:
\begin{itemize}
    \item \textit{click($a_i$)}: Clicking on element $a_i$
    \item \textit{type($a_i$, $i_i$)}: Typing input $i_i$ into element $a_i$
    \item \textit{select($a_i$, $i_i$)}: Selecting the option indexed by $i_i$ in element $a_i$
\end{itemize}

For the \textit{type} action, $i_i$ represents the textual input, whereas for the \textit{select} action, $i_i$ refers to the index of the option to be selected. The challenge lies in choosing the right action from this space to move closer to the task goal at each step, while considering the context of the current state and the entire task trajectory.

\begin{figure*}[h]
	\centering
        \includegraphics[width=\textwidth]{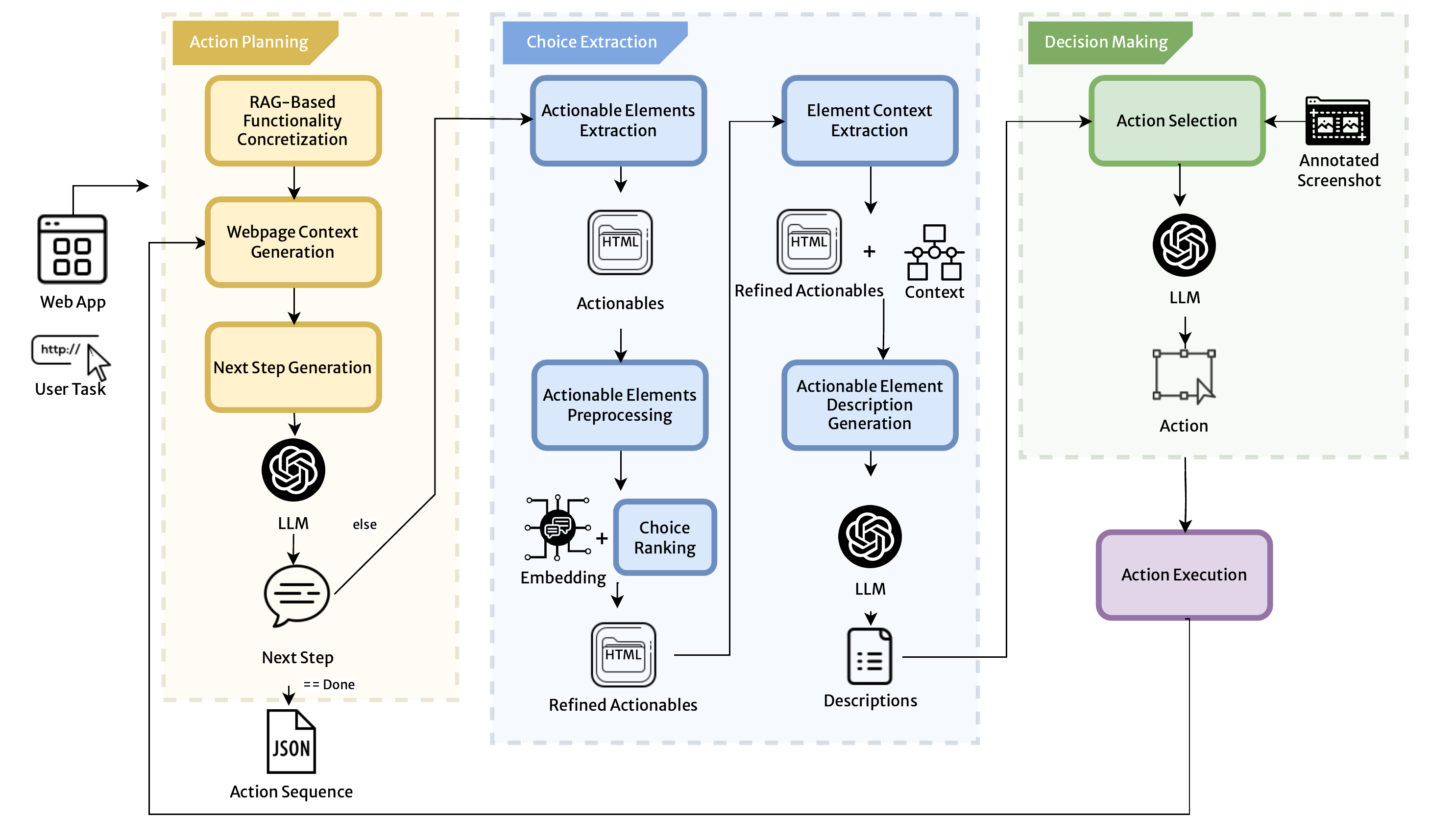}
	\vspace{-2em}
	\caption{Overview of \toolname.}
	\label{fig:overview}
\end{figure*}

To address the challenges outlined in previous sections, we introduce \toolname, which approaches the web application navigation problem as a \textit{sequential decision-making process}, where at each state $S_i$, it selects the optimal action $a_i$ from the available action space with the goal of completing the task $U$ efficiently by minimizing unnecessary actions and ensuring that all required steps are executed accurately. By integrating multi-modal inputs such as text and images, and combining task abstraction with dynamic action planning and selection, \toolname navigates complex web applications while adapting to diverse user functionalities. The input to \toolname consists of \emph{a one-sentence task or functionality description}, such as the example provided in the motivating example, along with the name of the intended website, and the output is \emph{a sequence of actions} to explore that task or functionality. \autoref{fig:overview} provides an overview of the proposed approach. \toolname employs a multi-phase, multi-model prompting methodology along with an option scoring system for web application navigation tasks, divided into three phases: \textbf{Action Planning}, \textbf{Choice Extraction}, and \textbf{Decision Making}.

\subsection{Action Planning}
The first phase of \toolname, as depicted in \autoref{fig:overview}, is Action Planning. In this phase, we first aim to concretize the functionality descriptions (for the case of functionality exploration) to guide the LLM in generating required task parameters (defined in Definition \autoref{def:user-task}), followed by a webpage context extraction for enhancing the LLM's understanding of the high-level context of the current state, and a next step prediction, aimed at optimizing the available options for the next phase.

\subsubsection{Retrieval-Augmented Functionality Concretization}
\label{subsec:rag}
In our approach to handling abstracted tasks, which involve functionality descriptions, we introduce an additional step involving the training set of a reference dataset, Mind2Web-Live \cite{pan:mind2web-live:arxiv2024}, containing concrete task descriptions. We first generate functionality descriptions for every concrete task in the dataset, utilizing \gpt, to construct a reference training database, where pairs of original concrete tasks from the Mind2Web-Live training set are matched with their abstracted counterparts. For example, the task mentioned in \autoref{sec:motivation}, \example, would be paired with \absexample. To generate concrete task descriptions from functionalities, we apply a Retrieval-Augmented Generation (RAG) method, which is shown to be effective in different code-related tasks~\cite{cedar:icse2023, toufique:semantic-augmentation-of-llm-prompts:icse2024, api-completion}. For each given functionality description as an input to \toolname, we retrieve the three most similar samples from the reference training database. The similarity measurement is based on cosine similarity between the text embeddings of the abstracted task and the tasks in the reference dataset. By converting the text of each task into high-dimensional vectors, cosine similarity allows us to measure the angular distance between these vectors, effectively capturing the semantic closeness of the tasks. This ensures that the retrieved samples are not just similar in surface-level wording but are also semantically aligned with the abstracted task. These retrieved samples are then used to prompt the language model to generate the concrete equivalent of the given abstracted task. The retrieval process ensures that no training data is directly reused during the test phase. Although the reference dataset is derived from the training data, the test dataset comprises distinct samples, separate from those used in training.

The generation of concrete tasks serves to improve the success rate of the agent's navigation by providing clearer inputs. However, our method is designed with flexibility, allowing it to adapt to the task description as closely as possible when navigating the web application. It continuously predicts the most appropriate action at any given moment and can redirect itself as needed. Even in cases where the exact concrete version of the task is not available, the method still attempts to select the most similar possible actions to achieve the desired outcome. It is important to emphasize that generating a concrete task from an abstract one differs from working with a fully predefined concrete task, as the former involves predicting the closest actions based on abstract descriptions, rather than relying solely on a given explicit instruction, which eliminates the need for defining detailed task descriptions.

\subsubsection{Webpage Context Generation}
\label{subsec:context}
To enhance the LLM's understanding of the web application's state at each step, we implement a webpage context generation process. This process generates a structured and abstract overview of the current webpage. To extract this context, we provide the LLM with the meta tag description of the website, the context from the previous state, the action that led to the current state, and a screenshot of the current webpage. The prompt instructs the model to describe the webpage's context abstractly, focusing on the webpage's primary purpose and the webpage's sub-functionalities. An example of an LLM-generated context for the landing page of \emph{Uniqlo} (the first state of our motivating example) is shown in \autoref{lst:webpage}. By generating this abstracted context for each state, we give the LLM a clearer understanding of the web application's flow, thereby improving the accuracy and relevance of its responses as it navigates through the web application.

\begin{lstlisting} [caption={Example of a generated contextual representation for a webpage},label={lst:webpage}]
{
	"context": "This page is the landing page for an online clothing store.",
	"sub_functionalities": [
		"users can browse different categories of clothing for various age groups",
		"users can view featured products and promotions",
		"users can navigate to other sections such as help and store locator",
		"users can switch between different languages",
		"users can search for specific items",
		"users can access their account, wishlist, and shopping cart"
	]
}
\end{lstlisting}

\subsubsection{Next Step Generation}
\label{subsec:planning}

We generate a prediction of the next step, $N_i$, based on the task description, history of actions, and the current webpage context (\autoref{subsec:context}). The model outputs either \emph{``Done''}, indicating task completion and ending navigation, or a sentence describing the next step. In the motivating example, the initial prompt could return \emph{``Click on the `Clothing' category''} or \emph{``Type `Blazer' in the search bar,''} as tasks can have multiple valid paths. In the motivating example, the user performs the second option.

The predicted next step description serves as a guide for the LLM to select the most appropriate actionable items and inputs for the action. Additionally, incorporating the next step helps to sort and refine the list of actionables based on their semantic similarity to the generated next step. This process enhances the accuracy and relevance of the subsequent actions, ensuring that the LLM can effectively navigate and interact with the web application.

\subsection{Choice Extraction}
As shown in \autoref{fig:overview}, the next phase, Choice Extraction, aims to prepare a list of options for the tool to choose from in the next phase. It involves extraction, preprocessing, refinement, context generation, and finally description generation of actionable elements.
To extract actionable elements, we utilize the tag names and event listeners of web elements in the current state to determine if they are considered actionable as defined in~\autoref{subsec:actionables}. For instance, in the motivating example discussed in \autoref{sec:motivation}, the actionable elements in the first state (the landing page) include, but are not limited to, all navigation buttons, the search button, and the carousel's navigation arrows.

\subsubsection{Actionable Elements' Preprocessing and Ranking}
\label{subsec:desc_gen}

Following the extraction of the actionable elements, we undertake a ranking process for optimization purposes. This process involves calculating the semantic similarity score ($score_{a_j}$) between the inner text of each actionable element ($a_j$) as defined in \autoref{eq:sim}, or the outer HTML of the element if no inner text exists, and the textual description of the next step ($N_i)$, which was generated in the initial workflow step. For this task, we employ cosine similarity of the embeddings of the two texts, denoted by $E_{a_j}$ and $E_{N_i}$, representing the textual embedding of the actionable element and the next step, respectively. 

\begin{equation}
\label{eq:sim}
score_{a_j} = \text{cosine\_similarity}(E_{a_j}, E_{N_i})
\end{equation}

If an actionable element has already been selected in previous steps, we reduce its score by half. This adjustment decreases the likelihood of the element being selected repeatedly by the LLM, thereby preventing repetitive selection and potential overemphasis on the same element. After this refinement step, we sort the elements by their scores (\emph{Choice Ranking}) and retain the top-K actionable elements, to form a list of \emph{Refined Actionables}. 

Moreover, we incorporate a preprocessing step to refine the HTML representation of actionable elements by removing certain attributes and child elements of \emph{style}, \emph{path}, and \emph{svg}. Additionally, we truncate excessively large HTML elements after the preprocessing step to optimize the data we pass to the LLM.

\subsubsection{Actionable Elements' Context Extraction}
\label{subsec:desc_gen}

In the next step, we extract contextual information for each element in the refined list of actionable elements and append it to its outer HTML. This involves identifying the visual neighbors of each element and incorporating their inner text. By calculating Euclidean distances within a defined threshold, we select the five closest elements to enhance the LLM's understanding of an element's context. For example, price or rating information near a product page link is included, providing context the link alone cannot offer. Additionally, available options for select elements are appended. An example is shown in \autoref{lst:actionable_json}, where a link to the Men’s section is paired with the neighboring text ``MEN,'' and the search input field with ``Search by keyword.''

\lstset{
    string=[s]{"}{"},
    stringstyle=\color{blue},
    comment=[l]{:},
    commentstyle=\color{black},
    tabsize=2,
    literate={\ \ }{{\ }}1
}
\begin{lstlisting} [caption={Example of an actionable element's contextual representation},label={lst:actionable_json},float]
{
    "0": {
        "outerHTML": "<a class='fr-global-nav-item px-s' href='/ca/en/men' ><div><span>men</span></div></a>",
        "neighbours": ["MEN"]
    },
    "1": {
        "outerHTML": "<input type="search" id="searchInput" autocomplete="off" class="fr-searchform-input" name="searchTerm">",
        "neighbours": ["Search by keyword"]
    }, ...
}
\end{lstlisting}

\subsubsection{Actionable Elements' Description Generation}
\label{subsec:desc_gen}

Subsequently, we prompt the LLM with the list of actionable elements and their contextual information to generate a one-sentence description for each element. We invoke the LLM to generate a description of the element's functionality, not its appearance or style, given the textual content of its visual neighbours. We perform this step in batches of 10 elements to optimize the process. An example of an output generated by the LLM for the example actionables shown in \autoref{lst:actionable_json} is shown in \autoref{lst:actionable_desc}. We use a cost-effective model, \gptthree, for this task, as it involves a high token count and is relatively simple for a language model to handle. For all other tasks, we use \gpt to take advantage of its advanced decision-making abilities. This step lays the foundation for the next phase, Decision Making, by providing a more comprehensive list of options to choose from.

\begin{lstlisting} [caption={Example of a generated description for an actionable element},label={lst:actionable_desc},float]
{
    "0": "A link to the men's section of the website.",
    "1": "An input field with search functionality",
    ...
}
\end{lstlisting}

\subsection{Decision Making}
The final phase of \toolname is Decision Making. In this phase, the tool selects the best action to move closer to completing the given task or functionality. As illustrated in \autoref{fig:overview}, this phase includes a core component that prompts the LLM with the available actions, accompanied by visual cues, to select the next action. Once the LLM selects an action, it is grounded back into the action space for execution.

\subsubsection{Action Selection and Input Generation}
\begin{figure}[htbp]
	\centering
        \includegraphics[width=0.7\textwidth]{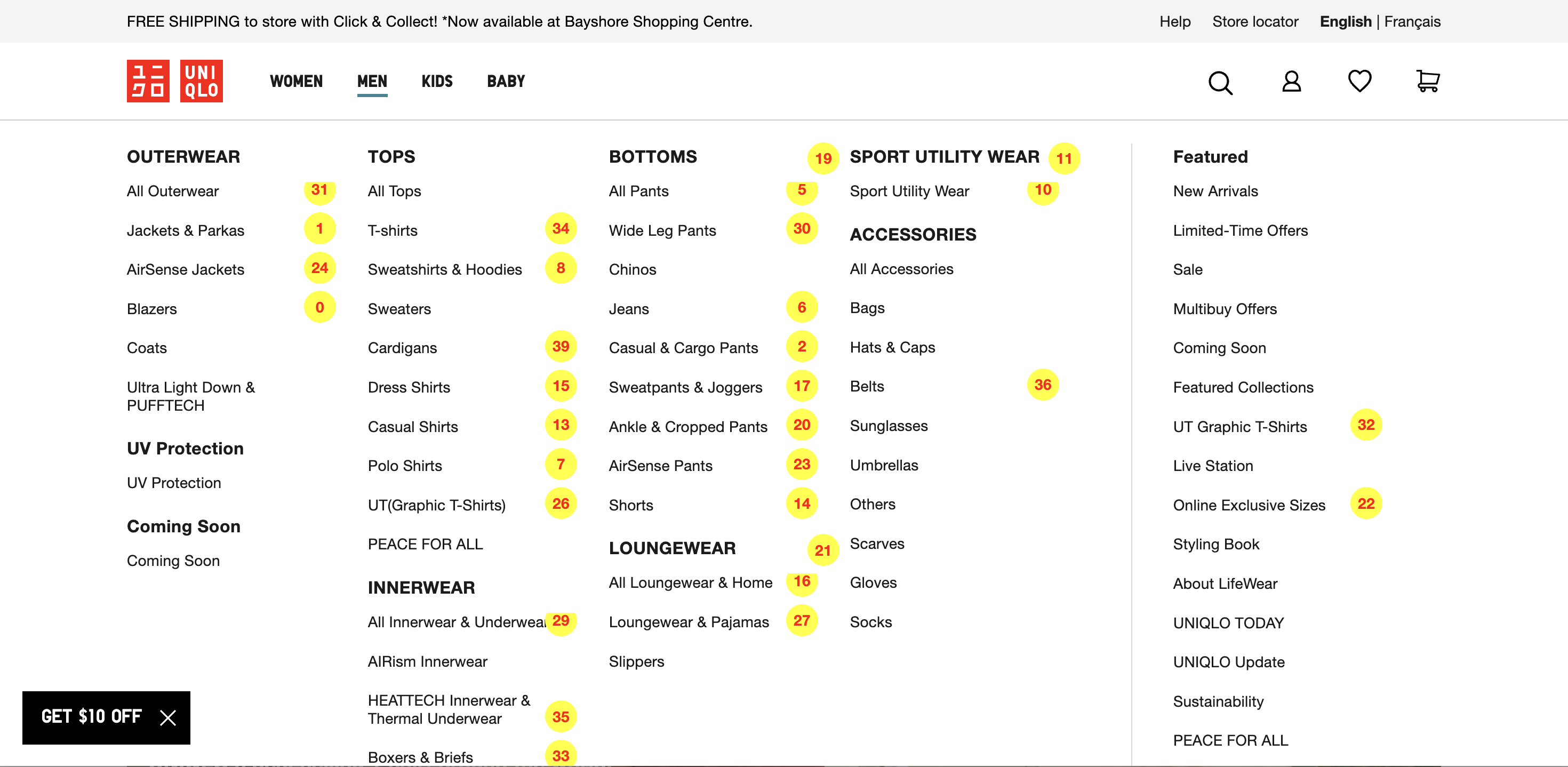}
	\vspace{0em}
	\caption{Example of screenshot annotation.}
	\label{fig:annotation}
\end{figure}

In this step, we first annotate a screenshot of the current webpage with visual cues indicating the index of each actionable element, thereby guiding the LLM regarding the visual representation and location of these elements. An example of an annotated screenshot is shown in \autoref{fig:annotation}. We sort the list of actionable elements based on the similarity score computed using~\autoref{eq:sim}. In the annotated screenshot example, the next step is to \emph{click on the Blazers category link}. As a result, the most semantically similar option is the link with "Blazers" as its text, which holds the 0 index in the list. By visually mapping the elements, the LLM can better interpret their layout and context, reducing ambiguity when determining the appropriate action. The graphical cues not only help the model associate descriptions with specific locations on the page but also allow it to consider additional visual attributes, such as the shape, colour, or size of the elements. These characteristics may influence the navigation, as certain actions could depend on recognizing visual attributes of an element. Incorporating these visual details enables the LLM to make more accurate predictions for interactions that rely on both spatial and graphical information, such as selecting the correct button or entering text into the right input field.

The LLM is then prompted with the annotated screenshot, the list of actionable elements' descriptions and their tag names, the history of previous actions, the task description, and the next step description to generate the next action to be executed on the web application. Based on this input, the LLM generates the next action to be performed on the web application. It selects from the given options and determines the type of action (as explained in ~\autoref{subsec:actionspace}) and required arguments. \toolname takes into account the need to click to activate the input field when processing \textit{type} actions. This prevents the generation of extra click actions, making the process more efficient. An example of the LLM's output for this step is provided in \autoref{lst:action_selection}. In this example, the LLM has selected the actionable element with an index of 0, which corresponds to a link to the ``Blazers'' section. This selection is then translated into executable Selenium commands, enabling interaction with the website. Finally, the selected action is executed via Selenium and appended to the action sequence for potential future reproduction.

\begin{lstlisting} [caption={A generated next action for \example},label={lst:action_selection},float]
{
    "index": 0,
    "action": "click"
}
\end{lstlisting}

The output of \toolname is a JSON file containing a list of actions for later reproduction. Each action includes the element's outer HTML, XPath, action type, and its arguments. An example sequence, shown in \autoref{lst:seq}, demonstrates \toolname performing the task \example by selecting links to the Men's section, the ``Blazers'' section, and a filter by color.

\begin{lstlisting} [caption={Generated action sequence for \example (truncated)},label={lst:seq},float]
    {
        "element": "<a class="fr-global-nav-item px-s" href="/ca/en/men" ><div><span>men</span></div></a>",
        "xpath": "/html/body/div[1]/..",
        "action": "click"
    },
    {
        "element": "<a data-category="navi" data-action="click" data-label="/men/outerwear/blazers" href="/ca/en/men/outerwear/blazers"><div><span class="inner">Blazers</span></div></a>",
        "xpath": "/html/body/div[1]/..",
        "action": "click"
    },
    {
        "element": "<a role="button" data-test="filter-by-colour"><span><span class="fr-accordion-head-text"><div>Color</div></span><span class="fr-accordion-head-arrow"><span>",
        "xpath": "/html/body/div[1]/..",
        "action": "click"
    } ...
}
\end{lstlisting}

\subsection{Stopping Criteria}
We define the task navigation stopping criteria as follows:
\begin{itemize}
    \item The output of the Next Step Generation step is \emph{``Done''}, indicating task completion as explained in~\autoref{subsec:planning}, or
    \item There are no available actions in the current state, or
    \item The trajectory length exceeds a predefined limit.
\end{itemize}

To avoid potential infinite loops and to control for scenarios where the agent may get stuck in repetitive or unproductive actions during navigation, we impose a limit of 20 steps per sequence. This step limit was chosen based on empirical analysis, as most tasks can be completed well within this range since the average reference task length in the \textit{Mind2Web-Live} dataset is \textit{7.9}. 

\subsection{Prompting}
We use zero-shot prompting to prevent overfitting and ensure the model generates solutions without relying on previously seen tasks, promoting generalization.

Our prompts are designed with a system prompt that outlines the task expectations, constraints, and required output format, providing the necessary context. The user prompt supplies specific inputs, including the task description, action history, and visual context. This structure ensures the LLM fully understands the task and has the relevant details to generate accurate, context-aware actions. 

% All prompts are detailed in Appendix \ref{prompts}.
% !TEX root =  ../main.tex
\section{Evaluation}
\label{sec:evaluation}
To assess the effectiveness of \toolname, we address the following research questions: 

\begin{itemize}    
    \item \textbf{RQ1}: How successful is \toolname in performing a web application navigation task?

    \item \textbf{RQ2}: How efficient is \toolname in performing a web application navigation task?

    \item \textbf{RQ3}: What role do different components play in the overall performance of \toolname?
\end{itemize}

\begin{table}[h!]
\centering
\scriptsize
\begin{tabular}{l|l|c}
\toprule
\textbf{Domain} & \textbf{Subdomain Name} & \textbf{Sample Count} \\ \hline
\multirow{5}{*}{Entertainment} 
& Event      & 5  \\  
& Game       & 3  \\  
& Movie      & 9  \\  
& Music      & 5  \\  
& Sports     & 9  \\  \hline
\multirow{6}{*}{Shopping}      
& Auto       & 7  \\  
& Department & 6  \\  
& Digital    & 6  \\  
& Fashion    & 3  \\  
& General    & 3  \\  
& Speciality & 13 \\ \hline
\multirow{6}{*}{Travel}        
& Airlines   & 5  \\  
& Car rental & 1  \\  
& Ground     & 9  \\  
& Hotel      & 3  \\  
& Other      & 11 \\  
& Restaurant & 6  \\ \hline
\multicolumn{2}{c|}{\textbf{Total}} & \textbf{104}  \\ 
\bottomrule
\end{tabular}
\caption{Distribution of samples by domain and subdomain in the Mind2Web-Live test set \cite{pan:mind2web-live:arxiv2024}.} 
\label{tab:dataset}
\end{table}

\subsection{Dataset Construction}
\label{subsec:dataset}
The importance of robust datasets for training and evaluating web agents and tools has spurred research into creating reliable resources. Notable datasets in the domain of web tasks include MiniWoB++ \cite{liu2018reinforcement:miniwoB:plus}, WebShop \cite{yao:webshop:neurips2022}, RUSS \cite{xu2021grounding}, and Mind2Web \cite{deng2024mind2web}. In particular, Mind2Web offers a rich collection of real-world user interactions with web applications, comprising task descriptions and corresponding action sequences. Its subset, Mind2Web-Live \cite{pan:mind2web-live:arxiv2024}, further refines this resource by excluding time-sensitive or obsolete tasks.

We utilize the test split of the Mind2Web-Live dataset, which comprises 104 samples, each outlining a specific user-defined task to be executed on a designated website. The websites are in three different domains, and 17 different subdomains, as shown in \autoref{tab:dataset}, with \textit{Movie} having the highest number of samples in \textit{Entertainment}, \textit{Speciality} having the most samples in \textit{Shopping} at 13, and \textit{Other} having the highest count at 11 in \textit{Travel}.

To generalize each task, we abstract the descriptions by omitting task-specific details and user input information. This abstraction process is designed to generate functionality descriptions, which are more streamlined and less time-consuming to compose compared to detailed task descriptions. We employ \gpt to remove the task-specific details from the descriptions. We refer to this refined dataset as the \textit{Mind2Web-Live-Abstracted}.

For the motivating example discussed in \autoref{sec:motivation}, which is defined as \example, the LLM generated abstraction is \absexample.

\subsection{Implementation}
\toolname is implemented in Python. In our implementation, Selenium~\cite{selenium} serves as the backbone for automated web interactions, executing actions on web applications as directed by the workflow. Selenium is a widely-used tool for automating web browsers, enabling us to programmatically control the browser to perform tasks such as clicking buttons, filling out forms, and navigating between pages. In conjunction with Selenium, we leverage LLMs to dynamically generate and refine instructions based on the current state of the web application. For all experiments, we set the temperature parameter to 0 to ensure deterministic and well-defined answers from the LLM. The embeddings are generated using MiniLM~\cite{wang:minilm:neurips2020}. For replaying the action sequences, we use Playwright~\cite{playwright}, a browser automation framework that allows interaction with web elements and generates visual execution traces. 

\subsubsection{Language Models} 
\begin{itemize}
    \item \gpt, 2024: The \gpt, released by OpenAI in 2024, is engineered to provide enhanced accuracy and efficiency. We leverage this model for the more complex tasks of action planning and decision-making.
    \item \gptthree, 2024: \gptthree is a cost-effective model in the GPT-4o series, developed by OpenAI. We leverage this model in the simpler task of generating descriptions for web elements.

\end{itemize}

\subsubsection{LLM-only Baseline (\gpt)} 
We implement a baseline model, utilizing \gpt for decision making. We only prompt the model once at each step to select an action. The model is provided with the HTML representation of actionable elements, along with the task description, to guide its action selection.

\subsection{Evaluation metrics}
\label{subsec:evaluation-metrics}

\header{Success Rate (\%)} 
In the context of our study, the success rate is defined as the proportion of user tasks that the agent successfully completes on a website. Specifically, a task is considered successfully completed if the agent achieves the intended outcome. The success rate is calculated by dividing the number of successfully completed tasks by the total number of tasks assigned to the agent and then multiplying by 100 to express it as a percentage. This metric has been used in the literature for assessing the performance of web agents \cite{zhou2024webarena, pan:mind2web-live:arxiv2024}.

\header{Trajectory Optimization Score (TOS)}
Trajectory length is defined as the number of steps or actions required to complete a given task, representing the length of the generated trajectory. Each step within the trajectory corresponds to an individual action executed by the agent.
We introduce the Trajectory Optimization Score as a metric for evaluating an agent’s performance in relation to the optimal solution. The score is calculated as the ratio of the ground truth trajectory length to the agent’s generated trajectory length for each task in the dataset. A value of 1 denotes that the agent’s trajectory aligns perfectly with the optimal path, while scores below 1 indicate inefficiencies or deviations from the ideal trajectory. To provide a comprehensive assessment, the average trajectory optimization score is computed across all tasks. For tasks in which the agent fails, a score of 0 is assigned, as higher scores correspond to better performance. This metric underscores the agent’s ability to optimize task completion.

\subsection{Human Evaluation}
For replaying the action sequences (trajectories), we utilize Playwright \cite{playwright}. Specifically, we generate a Playwright test script for each action sequence to capture execution traces in an interactive format supported by Playwright. These traces are subsequently employed for human evaluation. The authors review these traces, answering either ``Yes'' or ``No'' to assess whether the task was executed correctly. Each human evaluator (author) receives a set of execution traces, accompanied by step-by-step screenshots and a JSON file detailing the selected element (including its HTML representation and XPath), the corresponding action, and potential input to the action selected by \toolname. Following a provided rubric, each author independently examines the action sequence and determines whether the task was successfully completed. The results will be considered correct only if all authors reach a positive agreement. 

The Mind2Web-Live dataset includes human annotations \cite{pan:mind2web-live:arxiv2024}, which capture the sequence of human-performed actions for each task. In their paper, WebCanvas is evaluated against these human annotations. However, as previously noted in Section \ref{sec:motivation}, multiple valid approaches may exist for completing a task. Moreover, behavior cloning from expert demonstrations is prone to fragility and does not scale well \cite{gur2021adversarial}. In addition, annotations for functionalities are not available, and having a reference annotation for them to calculate the success rate is impractical because each functionality can correspond to multiple concrete tasks rather than a single specific task. We use the reference task length of the corresponding concrete task from the Mind2Web-Live dataset as a proxy for calculating the trajectory optimization score, with the reference task length representing one potential task aligned with the corresponding functionality.

To address this, we avoid evaluating our method against a single user-selected path. Instead, we follow a clearly defined rubric for manual evaluation. According to our rubric, a task is deemed successful if the intended functionality is explored and the correct task parameters are chosen. If additional steps are included in the sequence that do not alter the intended functionality—such as navigating to the website's homepage after completing the task without leaving the website—we still classify the task as successful. However, these additional steps contribute to the overall trajectory length, negatively impacting the trajectory optimization score.

\subsection{Effectiveness and Efficiency of \toolname (RQ1 \& RQ2)}
\label{ssec:effectiveprompt}
We evaluate our tool on both datasets: the original Mind2Web-Live dataset \cite{pan:mind2web-live:arxiv2024} and the abstracted dataset discussed in Section \ref{subsec:dataset}. We run the evaluation twice and report the average metrics of these two rounds. We report the metrics for both concrete and abstracted tasks independently.

\subsubsection{Comparison with State-of-the-Art Tools}
\label{ssec:comparison}
We compare our work with the WebCanvas web navigation agent introduced in \cite{pan:mind2web-live:arxiv2024}, as this is the only existing work to our knowledge that is evaluated on the Mind2Web-Live dataset. It is worth to mention that we use the updated WebCanvas agent, which utilizes \gptthree. 

Given that some tasks in the original Mind2Web dataset \cite{deng2024mind2web} have become unexecutable, we focus our evaluation on the refined dataset, Mind2Web-Live. To the best of our knowledge, no existing approaches specifically aim to address abstract functionalities rather than concrete tasks, such as those found in the \textit{Mind2Web-Live-Abstracted} dataset. Furthermore, no comparable datasets are currently available for such evaluation.

\begin{table*}[h]
    \label{tab:results}
    \centering    
    \caption{
    Web task navigation performance on the Mind2Web-Live and Mind2Web-Live-Abstracted datasets.}
    \scriptsize
    \begin{tabular}{c|cc|cc}
    \toprule
        \multirow{2}{*}{\textbf{Method}} & \multicolumn{2}{c|}{\textbf{Mind2Web-Live \cite{pan:mind2web-live:arxiv2024}}} & \multicolumn{2}{c}{\textbf{Mind2Web-Live-Abstracted}} \\
        & \textbf{SR (\%)} & \textbf{TOS} & \textbf{SR (\%)} & \textbf{TOS} \\
        \midrule
        WebCanvas \cite{pan:mind2web-live:arxiv2024} & \webcanvasconcretesr & \webcanvasconcretetos & \webcanvasabstractsr & \webcanvasabstracttos \\
        GPT-4o & \gptconcretesr & \gptconcretetos & \gptabstractsr & \gptabstracttos \\
        \toolname & \textbf{\oursconcretesr} & \textbf{\oursconcretetos} & \textbf{\oursabstractsr} & \textbf{\oursabstracttos} \\ 
        \bottomrule
    \end{tabular}
\end{table*}

For concrete tasks in the Mind2Web-Live dataset, \toolname achieves the highest SR of \oursconcretesr\%, significantly outperforming WebCanvas (\webcanvasconcretesr\%) and GPT-4o (\gptconcretesr\%). This indicates that \toolname can more effectively complete the user tasks defined in the dataset.
In terms of TOS, \toolname also outperforms WebCanvas, achieving a score of \oursconcretetos compared to WebCanvas' \webcanvasconcretetos. GPT-4o performs poorly with a TOS of \gptconcretetos, indicating it struggles with task optimization.

Similarly, for functionalities defined as abstracted tasks, \toolname shows a significantly better performance, with an SR of \oursabstractsr\% and a TOS of \oursabstracttos, outperforming both WebCanvas (SR: \webcanvasabstractsr\%, TOS: \webcanvasabstracttos) and GPT-4o (SR: \gptabstractsr\%, TOS: \gptabstracttos).
The gap in TOS between \toolname and WebCanvas is especially significant on this abstracted dataset, highlighting \toolname's ability to optimize navigation paths even when the tasks are more abstracted or generalized.

The specific tasks completed successfully differ between the abstracted and concrete datasets for the \gpt baseline, but the overall performance results remain consistent across both. The results demonstrate that relying solely on LLMs without any planning, context extraction, or additional guidance leads to poor navigation performance. The model struggles to consistently select the correct actions, as it lacks a structured approach to understanding the broader task context and sequence of steps.

\subsubsection{Effectiveness across domains}
The evaluation results of \toolname's performance on \textit{Mind2Web-Live} reveal notable differences in task success at both the website and subdomain levels. The Entertainment domain achieved the highest success rate at 48.4\%, and the Movie subdomain reached 88.9\%. Additionally, \toolname achieved a 100\% success rate on certain websites, demonstrating its capability to successfully interact and navigate those sites. However, some websites showed a 0\% success rate, highlighting instances where \toolname was unable to complete the tasks.

\begin{figure}[htbp]
	\centering
        \includegraphics[width=0.7\textwidth]{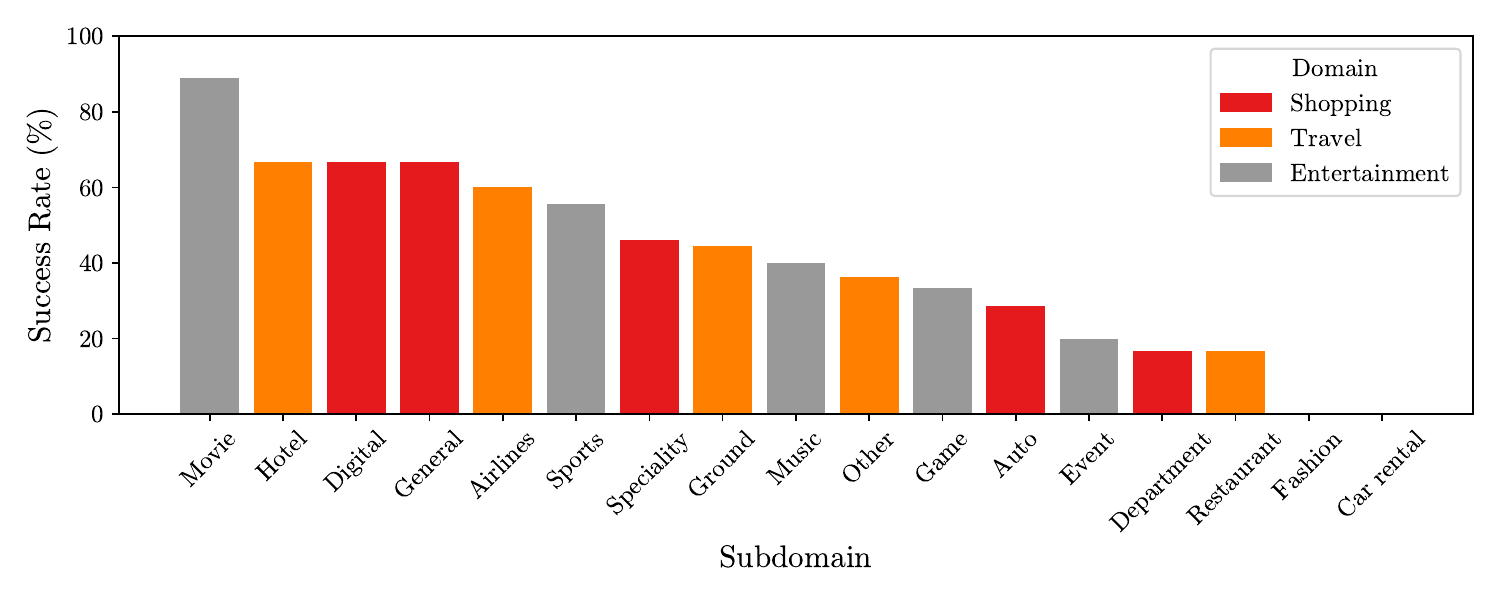}
	\vspace{-1.2em}
	\caption{Task success rate of \toolname for different subdomains on the Mind2Web-Live test set.}
	\label{fig:barchart-subdomain}
\end{figure}

\begin{figure}[htbp]
	\centering
        \includegraphics[width=\textwidth]{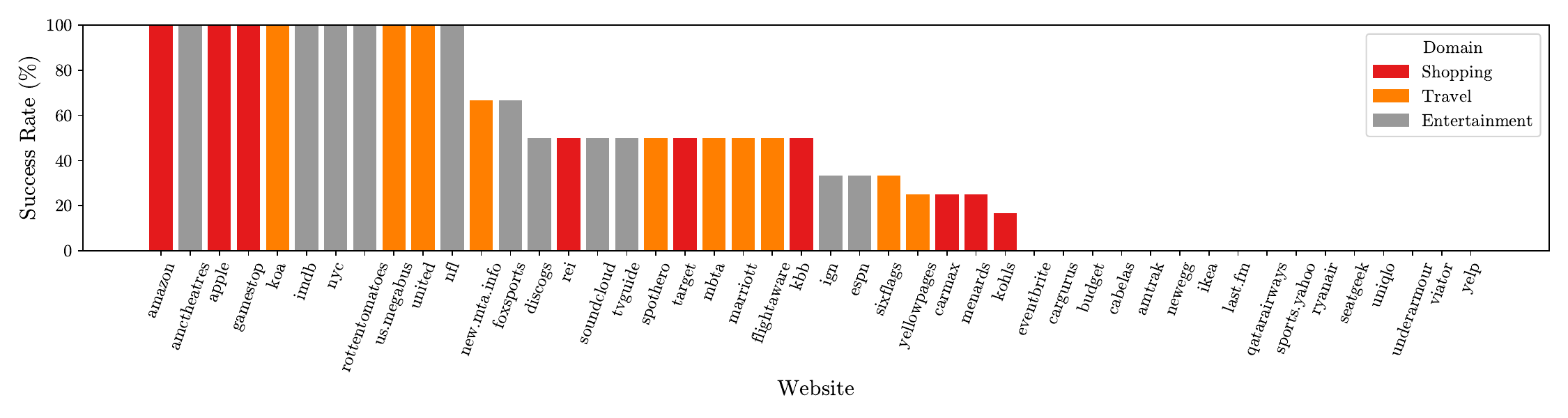}
	\vspace{-1.2em}
	\caption{Task success rate of \toolname for different websites on the Mind2Web-Live test set.}
	\label{fig:barchart-website} 
\end{figure}

\subsection{Ablation Study (RQ3)}

\header{Setup}
In our ablation study, we evaluate the impact of different variations of our method to better understand the contribution of each component. Specifically, we explore three key variations: 

\begin{itemize}
    \item First, a version of our method that omits the element description generation step.
    \item Second, a version of our method where the agent is provided with a greater number of actionable elements (top-K) as decision-making options during action selection, compared to a version with fewer actionable elements.
    \item Third, a version of our method that omits the context generation and action planning steps.
\end{itemize}

By examining these variations, we aim to identify the specific components of our approach that most significantly influence the agent's performance in navigating the web application. We select 20 random samples out of the 104 samples from the Mind2Web-Live test set for our ablation study. The rationale for selecting a smaller subset is due to the extensive number of variations in our ablation study, making it costly to run experiments across the entire dataset.

\header{Results}
In the first variant, we omit the element description generation (Section \ref{subsec:desc_gen}). Instead of prompting the LLM to generate descriptions of actionable elements, we directly use the textual content of these elements. We then provide a list of these textual contents, representing the actionable elements from which the agent can choose to interact in the next step of the approach. This modification leads to a decrease in the success rate, as shown in \autoref{tab:ablation}. The success rate of the original approach on the same test subset is 55\%, while the success rate of the approach without element description generation decreases to 50.0\%, and the TOS slightly decreases by 2\%, from 0.91 to 0.89.

In the second variant, we assess the effect of the number of actionable choices (K) for the action selection. We build this variant based on the previous variant but we increase the number of actionable elements to 100 from 40. We observe no difference in the success rate but a 9\% decrease in TOS. Therefore, we conclude that increasing the number of actionable choices does not improve performance. The number 40 was determined empirically and optimized for token efficiency.

In the third variant, we omit the webpage context generation (\autoref{subsec:context}) and action planning (\autoref{subsec:planning}). This results in a decreased success rate of 45\% compared to the original success rate of 55\%. Additionally, the TOS declines sharply by 54\%, reflecting a significant decrease in efficiency.

\begin{table*}[h]
    \centering
    \scriptsize
    \begin{tabular}{c|c|c}
    \hline
        \textbf{Method} & \textbf{SR (\%)} & \textbf{TOS}\\
        \hline
        \toolname & 55.0 & 0.91\\
        Without element description (40 elements) & 50.0 & 0.89\\
        Without element description (100 elements) & 50.0 & 0.81 \\
        Without planning & 45.0 & 0.42 \\ \hline
    \end{tabular}
    \caption{Performance on Mind2Web-Live subset, evaluating the role of different components.}
    \label{tab:ablation}
\end{table*}
% !TEX root =  ../main.tex
\section{Discussion}
The results of \toolname demonstrate the effectiveness of its multi-step LLM question and answering workflow in performing web application navigation tasks.

\header{Effectiveness of Multi-Modal Prompting} One key finding is the effectiveness of multi-modal prompting, which allows the system to combine textual descriptions with visual inputs to improve action selection. This approach reduces ambiguities often encountered in traditional web navigation agents that rely solely on textual inputs. By incorporating annotated screenshots and global and local contexts, \toolname can generate more accurate predictions of the next required action, leading to a higher task success rate compared to state-of-the-art tools like WebCanvas. Initially, we experimented with a variation of the method that excluded visual inputs (i.e., without screenshots) and observed a notable improvement when visual inputs were incorporated. This demonstrates that grounding the model in visual representations significantly enhances its predictive accuracy.

\header{Comparison with Baselines} 
Our experimental results demonstrate that \toolname achieves success rates of \oursconcretesr\% and \oursabstractsr\% on the Mind2Web-Live and Mind2Web-Live-Abstracted datasets, respectively. This represents a significant improvement over the WebCanvas agent, which attained success rates of \webcanvasconcretesr\% and \webcanvasabstractsr\% on the same datasets. Furthermore, \toolname's superior performance on the Mind2Web-Live-Abstracted dataset highlights its ability to effectively explore functionalities.

In many failed tasks, the WebCanvas agent would either navigate to the wrong website or move outside the correct website while attempting to complete the assigned task. In contrast, \toolname remains confined within the intended website and avoids such navigation errors.

The LLM-only baseline, utilizing \gpt, performs notably worse, achieving success rate of only \gptconcretesr\% on both the Mind2Web-Live and Mind2Web-Live-Abstracted datasets. In many failed cases, it repeatedly selects the same actionable element or chooses an incorrect action type for the element (e.g., using \emph{type} for a link element).

Furthermore, \toolname's lower success rate in exploring functionalities compared to concrete tasks may be due to the increased difficulty of generating inputs for functionalities, as specific task parameters are not provided.

\header{Impact of Different Components} The action planning and the refinement of actionable elements through context extraction and description generation was also found to contribute significantly to \toolname's performance. The ablation study revealed that omitting this step led to a reduction in task success rates, emphasizing the importance of this refinement process in ensuring accurate action generation. Interestingly, while increasing the number of actionable elements did not improve success rates, it did result in longer trajectories, indicating that having more decision options can introduce inefficiencies.

\header{Analysis of \toolname's Navigation Success Across Websites} Websites with higher success rates indicate that \toolname is more effective at navigating and completing tasks on these sites, likely benefiting from clearer structures or simpler interfaces. In contrast, lower success rates suggest that \toolname faced challenges, such as complex layouts or technical barriers, which impeded task completion. Some websites, such as \textit{viator}, required Captchas, preventing interaction, leading to a 0\% success rate. Other task failures were due to Webdriver issues, incorrect actionable choices, or improper input generation by the model. In several cases, \toolname came close to completing tasks but failed to finish them correctly, such as selecting a rating of 4 instead of 5 at the final step of a task requiring to find ``highly rated'' items. Action accuracy could potentially be improved by incorporating a more extensive feedback system that detects changes in the web application in real time.

\header{Future Directions} This study presents multiple opportunities for future research. One potential improvement is the implementation of an intermediate reward system, providing the model with feedback on partial task completion. This system could also be used as a new evaluation metric to improve performance measurement.

Another potential improvement could be to develop a more sophisticated feedback system, enabling the LLM to continuously monitor the effects of its actions and make more informed decisions based on web application state changes.

Moreover, a potential future direction could be the integration of a task or functionality description generation tool, such as the technique proposed by Alian et al. \cite{alian2024featurebasedapproachgeneratingcomprehensive}, which infers functionality descriptions for web applications. Such tools could be employed as a preliminary step to automatically generate functionality descriptions for \toolname. Once generated, \toolname can leverage these descriptions to perform a sequence of actions that explore the specified functionality. By incorporating this preliminary step, task descriptions could be generated automatically, leading to a more comprehensive and automated system for web application testing.

In summary, our work demonstrates the effectiveness of using LLMs for automated web application navigation and testing. Our tool, \toolname, can also function as a virtual assistant for impaired users, enhancing web accessibility.

\header{Threats To Validity}
A potential threat to validity is the risk of LLMs attempting to mimic the examples they have seen, rather than genuinely understanding and solving the tasks. This could result in inflated performance during evaluation, as the model may simply replicate patterns observed in the training data rather than adapting to new, unseen tasks. To mitigate this, we used different datasets for RAG and evaluation, ensuring that the model was not exposed to evaluation-specific data beforehand. Additionally, we employed zero-shot prompting, which prevents the LLM from relying on previously seen samples and encourages it to generate solutions based on task descriptions alone.

A key aspect of our approach involves abstracting tasks from the Mind2Web-Live dataset. However, this abstraction process may introduce errors or inconsistencies that could impact task execution. If the abstraction is too vague, it may result in task failures. To mitigate this, we use GPT-4o to remove task-specific details from the descriptions while ensuring that the task representations remain sufficiently generalized.

The dynamic and constantly evolving nature of web applications introduces a threat to validity, as web structures, elements, and behaviors may change after the evaluation is conducted. Tasks that are executable today may become unexecutable in the future due to these changes. To address this, we rely on the Mind2Web-Live dataset, curated to reflect up-to-date web tasks.
\section{Related Work}

\header{Web Crawling and Testing} Web application testing~\cite{balsam:web-application-testing-challenges:jss24} has advanced through automated approaches such as web crawling~\cite{crawljax}, enabling dynamic exploration and test generation~\cite{mesbah:invariant-based--testing:tse-12, biagiola:dependency-aware-web-test-generation:icst20}. A foundational contribution in this area is Crawljax~\cite{crawljax}, which explores dynamic web applications by triggering actions and analyzing the DOM. Recent advancements, such as QExplore\cite{sherin:explorer:sciencedirect23}, apply Q-learning to improve the exploration of web applications, while Yandrapally et al.\cite{yandrapally:near-duplicate:icse20} address near-duplicate states detection in web applications, enhancing model accuracy and testing efficiency by eliminating redundant states. Similarly, WebEmbed~\cite{stocco:neural-embedding:arxiv23} uses neural embeddings to abstract web page states, improving duplicate detection and coverage during web crawling. Overall, these approaches focus on maximizing exploration by interacting with user interfaces to generate new states. In contrast, \toolname adopts a functionality-guided approach to web application exploration, navigating based on a given functionality or task. This strategy allows for more structured and goal-oriented navigation, enhancing the ability to complete complex functionalities within the application.

\header{LLM-based Mobile and Web Testing}
Recent studies have increasingly focused on using LLMs to automate various aspects of web testing. Advances in LLMs have enabled their application in a range of testing domains, including GUI testing for both web and mobile applications~\cite{liu2024make, liu2024testing, fillintheblank}. Additionally, LLMs are being used to generate automated web form tests, simulating user interactions to validate form functionalities~\cite{li2024leveraging, alian2024bridging}.

\header{Web Agents}
In recent years, there has been increasing interest in developing web agents capable of determining and executing sequences of actions within web applications based on natural language instructions~\cite{li2021glider, mazumder2020flin, xu2021grounding, gur2018learning}. The key challenge lies in creating agents that can interpret complex user instructions and translate them into appropriate web-based actions.

The advent of language models has introduced a new paradigm for web agents, removing the need for expert demonstrations~\cite{gur2018learning}, human-designed heuristics~\cite{zheng2021automatic}, or restricting actions to predefined sets~\cite{xu2021grounding}. These models leverage advanced reasoning capabilities to determine the next steps based on natural language inputs~\cite{Wang2023ASO, gur2022understanding, furuta2023multimodal, sodhi2023heap}. This flexibility allows LLM-based agents to handle a wider range of web tasks, making them more versatile than previous methods.

Recent work, such as WebCanvas~\cite{pan:mind2web-live:arxiv2024}, uses LLMs to guide agents in navigating web environments. However, it relies heavily on detailed task descriptions and lacks support for broad functionalities, limiting its applications to complex tasks. In contrast, our approach supports a wide range of functionalities without requiring detailed task descriptions.

\header{LLMs for Software Testing} 
LLMs are increasingly applied in software testing, including unit test generation~\cite{xie:chatunitest:arxiv23, yuan:evaluate-chatgpt:arxiv2023, cedar:icse2023}, security testing~\cite{zhang2023well}, and defect reproduction~\cite{kang2023large}. Additionally, studies have explored their role in predicting flaky tests~\cite{fatima2022flakify} and automating the generation and migration of test scripts~\cite{yu2023llm}.
\section{Conclusion}
In conclusion, this paper introduces \toolname, a functionality-guided web application navigation tool that leverages large language models to enhance automated web testing. By framing web task navigation as a question-and-answer task and utilizing a three-phase approach with a combination of advanced and cost-efficient models, \toolname significantly improves task success rates and testing efficiency. Evaluated on the Mind2Web-Live dataset, \toolname demonstrated substantial improvements over the state-of-the-art, achieving a success rate of \oursconcretesr\% on user tasks and \oursabstractsr\% on functionalities. These results represent an improvement of \comparisonconcrete\% and \comparisonabstract\% over the WebCanvas agent, along with a trajectory optimization score increase of \comparisonconcretetos\% and \comparisonabstracttos\%. \toolname’s integration of multi-modal inputs and functionality-driven navigation enables it to navigate complex web tasks effectively. These contributions lay the groundwork for advancing automated web testing tools.

\section{Data Availability}
The implementation of our technique, \toolname, along with the accompanying Mind2Web-Live-Abstracted dataset, has been made publicly available~\cite{naviqate}. Additionally, comprehensive documentation for \toolname is provided to ensure the reproducibility of our findings.

\bibliographystyle{ACM-Reference-Format}
\interlinepenalty=10000
\pagebreak
\bibliography{references}

\pagebreak

\end{document}